\documentclass[12pt]{article}
\begin{document}
\pagestyle{plain}
\setcounter{page}{1}
\begin{center}
{\large\bf Bimetric Gravity Theory, Varying Speed of Light and the Dimming of
Supernovae}
\vskip 0.3 true in {\large J. W. Moffat}
\vskip 0.3 true in
{\it Department of Physics, University of Toronto, Toronto,
Ontario M5S 1A7, Canada}
\vskip 0.3 true in
and
\vskip 0.3 true in
{\it Perimeter Institute for Theoretical Physics, Waterloo, Ontario N2J
2W9, Canada}

\vskip 0.3 true in

\end{center}
\begin{abstract}%
In the bimetric scalar-tensor gravitational theory there are two frames associated
with the two metrics ${\hat g}_{\mu\nu}$ and $g_{\mu\nu}$, which are linked
by the gradients of a scalar field $\phi$. The choice of a comoving frame for the
metric ${\hat g}_{\mu\nu}$ or $g_{\mu\nu}$ has fundamental physical consequences for
local observers in either metric spacetimes, while maintaining diffeomorphism
invariance. When the metric $g_{\mu\nu}$ is chosen to be associated with comoving
coordinates, then the speed of light varies in the frame with the metric ${\hat
g}_{\mu\nu}$. Observers in this frame see the dimming of supernovae because of the
increase of the luminosity distance versus red shift, due to an increasing speed of
light in the past universe. Moreover, in this frame the scalar field $\phi$
describes a dark energy component in the Friedmann equation for the cosmic scale
without acceleration. If we choose ${\hat g}_{\mu\nu}$ to be associated with
comoving coordinates, then an observer in the $g_{\mu\nu}$ metric frame will observe
the universe to be accelerating and the supernovae will appear to be farther away.
The theory predicts that the gravitational constant $G$ can vary in spacetime, while
the fine-structure constant $\alpha=e^2/\hbar c$ does not vary. The problem of
cosmological horizons as viewed in the two frames is discussed.
\end{abstract}


\vskip 0.3 true in
e-mail: john.moffat@utoronto.ca

\section{Introduction}

Einstein's special relativity is based on the two postulates:
\begin{enumerate}

\item All physical laws are invariant with respect to inertial reference frames;

\item The speed of light is a universal constant with respect to all inertial
reference frames.

\end{enumerate}

The first postulate states that physical phenomena have the same appearance in all
inertial reference frames. One obtains the same results for all measurable quantities
and clocks are synchronized by means of two-way speed of light measurements. The
second postulate states that there is no preferred reference frame and that
globally, physical laws appear the same in all inertial frames. For a
description of a consistent special relativity theory, it is sufficient to
just adopt postulate (1)~\cite{Selleri}. However, the speed of light is
still a constant but this constant can vary from one reference frame to
another. The spacetime coordinates are defined so that the one-way speed
of light is constant.

In a Galilean spacetime the physical existence of an absolute time is postulated. The
global concept of past, present and future is the same in all inertial
frames. Simultaneous events in spacetime can occur in all reference frames. There
exists a unique separation of past and future events. The speed of light is only
constant in the Newtonian rest frame, which is a preferred frame of reference.
Electromagnetic waves are limited not to exceed the speed of light.

The first postulate of Einstein's special relativity is well established
by {\it local} experiments, whereas the second one is not. In contrast to Galilean
relativity with its absolute time, simultaneity is not an absolute concept, but a
relative one, depending on the motion of the observer. However, the first postulate
of Einstein's special relativity may not be so well established at the scale of
cosmology or at short distances.

Although the local Lorentz invariance of the laws of physics is a
mathematical and physically attractive idea, there have been several
reasons to be concerned about its generality at a fundamental level. It
should be emphasized that there is no experiment that determines the
one-way speed of light, for this would suppose that we can synchronize
physical clocks by some way other than by finite speed of light signals.
The main reason to be concerned with the generality of Einstein's special
relativity is the conflict between the two cornerstones of modern physics,
General relativity (GR) and quantum mechanics.

In GR the concept of time disappears as a physically important
quantity, because of the reparameterization invariance of the theory. This has led to
considerable research in attempts to understand how to construct a consistent theory
of quantum gravity~\cite{Isham}. When we attempt to ``quantize'' spacetime,
we find it difficult to maintain our classical ideas about local causality
and special relativity associated with the Minkowski lightcone. Moreover,
in quantum mechanics time is an external parameter, whereas in special
relativity space and time are on an equal footing.

In cosmology, a comoving coordinate time does appear in the framework of a Friedmann,
Robertson and Walker (FRW) spacetime, in which the time $t$ appears as a universal
time which measures the lifetime of the universe. However, the FRW metric is subject
to a diffeomorphism reparameterization transformation to non-comoving coordinates and
the notion of a ``universal'' absolute time disappears.

The idea that a variable speed of light can solve the initial value problems of
standard big-bang cosmology has received considerable
attention~\cite{Moffat,Moffat2,Magueijo}. From our
considerations of relativity theory, we learn that it is not possible to have a
varying speed of light, without modifying Einstein's formulation of special
relativity and GR.

In our first treatment of a varying
speed of light theory~\cite{Moffat,Moffat2}, we introduced a formalism for
spontaneously breaking the local Lorentz invariance of the vacuum within GR. The
symmetry of the homogeneous Lorentz group $SO(3,1)$ was spontaneously broken to the
Galilean group $O(3)$ in the early universe. This introduced the concept of an
``absolute'' time and a formulation of quantum gravity that did not find itself in
conflict with the concept of time in quantum mechanics. As the universe expanded, a
phase transition occurred that reinstated the four-dimensional symmetry of
the homogenous Lorentz group. However, the speed of light was only allowed
to change discontinuously from one constant value near the beginning of the
universe to a much smaller value corresponding to the presently measured
value of the speed of light $c_0=299\, 792\, 458\,\,m/s$. A consistent
treatment of a varying speed of light within a gravitational theory was
lacking. In order to remedy this problem, a bimetric gravity theory (BGT) was
constructed, which provided a consistent diffeomorphism invariant formalism for
descibing varying speed of light
phenomena~\cite{ClaytonMoffat,ClaytonMoffat2,ClaytonMoffat3,ClaytonMoffat4,Drummond}.

In the following, we shall focus on the scalar-tensor bimetric
theory~\cite{ClaytonMoffat2,ClaytonMoffat4}. The theory can be the preliminary stage
for a quantum gravity theory that does not conflict with some of the basic properties
of quantum mechanics. The physical effects of having two different spacetime metrics
may be expected to be observationally detectable only at cosmological
scales, when dark matter and dark energy dominate the gravitational effects, or at
short distances. The BGT satisfies a matter conservation law
that assures that matter test particles travel along geodesics and do not violate
the weak equivalence principle, which is observationally well-tested~\cite{Will}.
Recently, the predictions of a primordial fluctuation spectrum were obtained from the
scalar-tensor bimetric gravity theory and compared to the CMB
data~\cite{ClaytonMoffat5}.

The observations of supernovae (SNe Ia) at red shifts $0.35 < z < 1.75$ have shown
that they appear to be farther away and therefore fainter than is to be expected
from the standard decelerating model of the universe~\cite{Perlmutter}. This
has led to the introduction of dark energy models which generate an accelerating
expansion of the universe, based on either a cosmological constant or a form of
quintessence~\cite{Peebles}. The major problem with these models is that they lead
to a coincidence problem, and a fine-tuned tiny mass smaller than the Hubble
parameter, $m_Q \sim 10^{-33}$ eV, and small couplings to visible matter
which must satisfy fifth-force constraints~\cite{Carroll}. Moreover, an
eternally accelerating universe will produce de Sitter event horizons that
cause problems for string theory and quantum gravity
theories~\cite{Fischler}.

Recently, an alternative explanation of the SNe Ia observations was proposed without
the acceleration caused by dark energy, based on the idea of flavor
oscillations~\cite{Terning}. An axion and a photon
mixing causes an attenuation of the photon flux from distant systems. The
model postulates an axion mass $m_A \sim 10^{-16}$ eV and a coupling with a mass
scale $M_S\sim 10^{11}$ GeV. A higher-dimensional model with leakage into
the higher dimension has also been proposed as an alternative to the
standard cosmological constant or quintessence models~\cite{Freese}.

In the following, we shall consider the problem of the dimming of supernovae
within the BGT in the variable speed of light (VSL) frame.

The Friedmann equation obtained from
the field equations of BGT in the VSL frame leads to dark matter and dark
energy components which give the experimentally observed result
$\Omega_{\rm Tot}\sim 1$, without generating an accelerating expansion of
the universe. The observed dimming of the
SNe Ia is affected by an increase of the red shift by 10\%-15\%, due to an increase in the
speed of light in the past, which increases the luminosity distance to the
supernovae.

We consider the consequences for other fundamental ``constants'' of nature
such as Newton's constant $G$ and the fine-structure constant
$\alpha=e^2/\hbar c$ and discuss the problem of cosmological horizons in
BGT.

\section{Bimetric Scalar-Tensor Gravity}

Our scalar-tensor bimetric gravity theory (BGT)
~\cite{ClaytonMoffat2,ClaytonMoffat4} is described by two metrics
called the ``matter metric'' ${\hat g}_{\mu\nu}$ and the ``gravitational
metric'' $g_{\mu\nu}$, respectively, connected by the ``biscalar'' field
$\phi$ through the equation
\begin{equation}
\label{eq:bimetric}
\hat{g}_{\mu\nu}=g_{\mu\nu}+B\partial_\mu\phi\partial_\nu\phi,
\end{equation}
where the constant $B$ has dimensions of $[{\rm length}]^2$ and is chosen to be
positive. The inverse metrics ${\hat g}^{\mu\nu}$ and $g^{\mu\nu}$ satisfy
\begin{equation}
\label{kronecker}
{\hat g}^{\mu\alpha}{\hat g}_{\nu\alpha}={\delta^\mu}_\nu,\quad
g^{\mu\alpha}g_{\nu\alpha}={\delta^\mu}_\nu.
\end{equation}

We shall restrict ourselves to the physical non-degenerate case:
${\rm Det}({\hat g}_{\mu\nu})\not= 0$.
The metric ${\hat g}_{\mu\nu}$ is used to construct the matter action and
can be said to be the geometry on which matter fields propagate. It is the
combination of the gravitational metric and the biscalar field that we
consider as being the gravitational fields of our theory. One of the
satisfactory features of the theory is that the formalism is generally
covariant (diffeomorphism invariant), which guarantees that basic
properties of the theoretical structure are consistent.

The model that we introduced in~\cite{ClaytonMoffat2,ClaytonMoffat4}
consisted in a self-gravitating scalar field coupled to matter
through the matter metric (\ref{eq:bimetric}), with the action
\begin{equation}
S=S_{\rm grav}+S_{\phi}+\hat{S}_{\rm M},
\end{equation}
where
\begin{equation}
S_{\rm grav}=-\frac{1}{\kappa}\int d\mu (R[g]+2\Lambda),
\end{equation}
$\kappa=16\pi G/c_0^4$, $\Lambda$ is the cosmological constant, and
we employ a metric with signature $(+,-,-,-)$.  We will write, for
example, $d\mu=d^4x\,\sqrt{-g}$ and $\mu=\sqrt{-g}$
for the metric density related to the gravitational metric
$g_{\mu\nu}$, and similar definitions of $d\hat{\mu}$ and
$\hat{\mu}$ in terms of the matter metric
$\hat{g}_{\mu\nu}$. The minimally-coupled scalar field
action is given by
\begin{equation}
S_{\rm \phi}=\frac{1}{\kappa}\int d\mu\,
\Bigl[\frac{1}{2}g^{\mu\nu}\partial_\mu\phi\partial_\nu\phi-V(\phi)\Bigr],
\end{equation}
where the scalar field $\phi$ has been chosen to be dimensionless. The
energy-momentum tensor for the scalar field that we will use is
given by
\begin{equation}
T^{\mu\nu}_\phi=\frac{1}{\kappa}\Bigl[
g^{\mu\alpha}g^{\nu\beta}\partial_\alpha\phi\partial_\beta\phi
-\frac{1}{2}g^{\mu\nu}g^{\alpha\beta}\partial_\alpha\phi\partial_\beta\phi
+g^{\mu\nu}V(\phi) \Bigr],
\end{equation}
and is the variation of the scalar field action with respect to
the gravitational metric:
\begin{equation}
\frac{\delta {S}_{\phi}}{\delta g_{\mu\nu}}
 =-\frac{1}{2}\mu T_\phi^{\mu\nu}.
 \end{equation}

We shall use the metric (\ref{eq:bimetric}) to construct the matter action
$\hat{S}_{\mathrm{M}}$,
resulting in the identification of $\hat{g}_{\mu\nu}$
as the metric that provides the arena on which matter fields interact.
The matter action $\hat{S}_{\mathrm{M}}[\psi^I] =
\hat{S}_{\mathrm{M}}[\hat{g},\psi^I]$, where $\psi^I$ represents all the
matter fields in spacetime. The
energy-momentum tensor
\begin{equation}
\label{eq:matterEM}
\frac{\delta {S}_{\mathrm{M}}}{\delta \hat{g}_{\mu\nu}}
 =-\frac{1}{2}\hat{\mu}\hat{T}^{\mu\nu},
\end{equation}
satisfies the conservation laws
\begin{equation}\label{eq:matterconservation}
\hat{\nabla}_\nu\Bigl[\hat{\mu}\hat{T}^{\mu\nu}\Bigr]=0.
\end{equation}
This follows as a consequence of the matter field equations
only~\cite{ClaytonMoffat2,ClaytonMoffat4}.
Here, $\hat{\nabla}_\mu$ is the
metric compatible covariant derivative determined by the matter
metric: $\hat{\nabla}_\alpha\hat{g}_{\mu\nu}=0$.

The gravitational
field equations are given by~\cite{ClaytonMoffat2,ClaytonMoffat4}:
\begin{equation}
\label{eq:Einsteins eqns}
G^{\mu\nu}=\Lambda g^{\mu\nu}
+\frac{\kappa}{2}T^{\mu\nu}_\phi
+\frac{\kappa}{2}\frac{\hat{\mu}}{\mu}\hat{T}^{\mu\nu}.
\end{equation}
The scalar field wave equation is
\begin{equation}
\label{eq:scalar FEQ}
\nabla_\mu\nabla^\mu\phi+V^\prime(\phi)-\kappa\frac{\hat\mu}{\mu}{\hat
T}^{\mu\nu}{\hat\nabla}_\mu{\hat\nabla}_\nu\phi=0,
\end{equation}
where $\nabla_\mu$ is the covariant derivative with respect to the metric
$g_{\mu\nu}$.

From the definition (\ref{eq:bimetric}), we obtain the inverses
\begin{equation}
{\hat g}^{\mu\nu}=g^{\mu\nu}-\frac{B}{I}\nabla^\mu\phi\nabla^\nu\phi,
\end{equation}
and
\begin{equation}
g^{\mu\nu}={\hat
g}^{\mu\nu}+\frac{B}{K}\hat\nabla^\mu\phi\hat\nabla^\nu\phi,
\end{equation}
where
\begin{equation}
I=1+Bg^{\mu\nu}\partial_\mu\phi\partial_\nu\phi,\quad K=1-B{\hat g}^{\mu\nu}
\partial_\mu\phi\partial_\nu\phi.
\end{equation}
It follows that $IK=1$ and we have defined $\nabla^\mu\phi=g^{\mu\nu}\partial_\nu\phi$
and
\begin{equation}
\hat\nabla^\mu\phi={\hat g}^{\mu\nu}\partial_\nu\phi=K\nabla^\mu\phi.
\end{equation}

We will assume a perfect fluid form for the matter
fields
\begin{equation}
 \hat{T}^{\mu\nu}=
 \Bigl(\rho+\frac{p}{c_0^2}\Bigr)\hat{u}^\mu\hat{u}^\nu
 -p\hat{g}^{\mu\nu},
\end{equation}
where
$\hat{g}_{\mu\nu}\hat{u}^\mu\hat{u}^\nu=1$ and ${\hat u}^\mu=dx^\mu/d{\hat s}$.

A fundamental feature of the BGT is that there is a separate frame associated
with each metric, and each frame or metric has its own light cone.
Because of the biscalar field linkage between the two frames, the two light cones
{\it cannot be physically identical}. If we choose a small patch of spacetime
in the ``gravitational'' frame with the metric $g_{\mu\nu}$, then we
can locally make the $g_{\mu\nu}$ metric equal the flat, Minkowski
metric, i.e. $g_{\mu\nu}=\eta_{\mu\nu}$ where $\eta_{\mu\nu}={\rm
diag}(1,-1,-1,-1)$. This yields
\begin{equation}
\label{Minkowski}
ds^2=\eta_{\mu\nu}dx^\mu dx^\nu.
\end{equation}
However, from (\ref{eq:bimetric}) we obtain
\begin{equation}
\label{nonMinkowski}
d{\hat s}^2=\eta_{\mu\nu}dx^\mu dx^\nu+B\partial_\mu\phi\partial_\nu\phi dx^\mu dx^\nu.
\end{equation}
If we perform a local Lorentz transformation
\begin{equation}
x^{'\mu}={\Lambda^\alpha}_\nu x^\nu,
\end{equation}
where ${\Lambda^\alpha}_\nu$ are constant tensor coefficients that satisfy
\begin{equation}
{\Lambda^\alpha}_\mu\Lambda_{\alpha\nu}=\eta_{\mu\nu},
\end{equation}
then (\ref{Minkowski}) and (\ref{nonMinkowski}) both remain form invariant. This
follows if we
have $V_\mu=\partial_\mu\phi$, where $V'_{\mu}={\Lambda_\mu}^\alpha V_\alpha$ under a
Lorentz transformation.
If the light cone equation $ds^2=0$ is satisfied, then $d{\hat s}^2\not= 0$ along the
same light cone unless  $\phi= 0$. However, $d{\hat s}^2=0$ along an expanding light
cone that encloses the light cone $ds^2=0$. We observe that whereas we can transform
away the Christoffel symbols $\Gamma^\lambda_{\mu\nu}$ at a spacetime point, we
cannot transform away the scalar field $\phi$ at this point, nor the Christoffel
symbol ${\hat\Gamma}^\lambda_{\mu\nu}$.

If we choose the vacuum expectation value ${\cal V}_\mu=<V_\mu>_0\not= 0$, then the
gradient of the scalar field $\phi$ acts as a spontaneous symmetry breaking field,
so that it is not possible to maintain the local Lorentz invariance of the vacuum
states of both ${\hat g}_{\mu\nu}$ and $g_{\mu\nu}$ metrics. Thus, a preferred frame
is chosen under the spontaneous symmetry breaking of the vacuum, since ${\cal V}_\mu$
picks out a preferred direction in spacetime. This frame can correspond to
a timelike vector ${\cal V}_\mu=({\cal V},0,0,0)$, leading to
a spontaneous breaking of the homogeneous Lorentz group
$SO(3,1)\rightarrow O(3)$ i.e. only the symmetry group of rotations would be
preserved. In GR there is only one rigid light cone and one metric and local Lorentz
invariance is strictly maintained.

In the BGT developed in earlier papers, we assumed that in the VSL
frame all particles move with the the speed of light
$c(t)$. We can generalize this theory, so that each particle has
its own light cone by adopting the metric definition:
\begin{equation}
{\hat g}_{I\mu\nu}=g_{\mu\nu}+B_I\partial_\mu\phi_I\partial_\nu\phi_I,
\end{equation}
where the label $I$ denotes the specific particle with its associated speed
$c_I(t)$.

\section{Bimetric Gravity Cosmology}

Let us now consider a cosmological scenario, imposing
homogeneity and isotropy on spacetime and writing the ``gravitational frame''
metric $g_{\mu\nu}$ in comoving form
\begin{equation}
ds^2=c_0^2dt^2-R^2(t)\biggl[\frac{dr^2}{1-kr^2}+r^2d\theta^2+r^2\sin^2\theta
d\phi^2\biggr],
\end{equation}  where we employ a dimensionless radial variable $r$
and $k=0,\pm 1$ for flat, closed and open hyperbolic spatial topologies,
respectively. Then, the metric in what we call the
VSL frame takes the form
\begin{equation}
\label{vslmetric}
d{\hat s}^2
=c^2(t)dt^2-R^2(t)\biggl[\frac{dr^2}{1-kr^2}+r^2d\theta^2+r^2\sin^2\theta
d\phi^2\biggr],
\end{equation}
where
\begin{equation}
\label{lightspeed}
c(t)=c_0I^{1/2},
\end{equation}
and
\begin{equation}
I=1+\frac{B}{c_0^2}\dot\phi^2.
\end{equation}
An overdot indicates a derivative with respect to the time variable
$t$.

The matter stress-energy tensor is
\begin{equation}
\hat{T}^{00}=\frac{\rho}{I},\quad
{\hat T}^{0i}=0,\quad \hat{T}^{ij}=\frac{p}{R^2}\gamma^{ij},
\end{equation}
where $\gamma^{ij}\,\,(i,j=1,2,3)$ denotes the spatial metric.
The conservation laws become
\begin{equation}\label{eq:cosm cons}
\dot{\rho}+3H\Bigl(\rho+\frac{p}{c_0^2}\Bigr)=0,
\end{equation}
where $H=\dot{R}/R$ is the Hubble function.

We see that in the VSL frame the speed of light depends on
time, whereas in the variable speed of gravitational waves (VSGW) frame, the speed of
light is constant and the speed of gravitational waves depends on time. The latter
frame is arrived at by writing the matter metric ${\hat g}_{\mu\nu}$ in comoving
form with constant $c_0$:
\begin{equation}
d{\hat s}^2
=c_0^2dt^2-R^2(t)\biggl[\frac{dr^2}{1-kr^2}+r^2d\theta^2+r^2\sin^2\theta
d\phi^2\biggr].
\end{equation}

We now obtain
\begin{equation}
ds^2=v^2_gdt^2-R^2(t)\biggl[\frac{dr^2}{1-kr^2}+r^2d\theta^2+r^2\sin^2\theta
d\phi^2\biggr],
\end{equation}
where $v_g$ is the speed of gravitational waves
\begin{equation}
v_g(t)=c_0K^{1/2},
\end{equation}
and
\begin{equation}
K=1-\frac{B}{c_0^2}\dot\phi^2.
\end{equation}
We see that the speed of gravitational waves $v_g$ in the VSGW frame is predicted for
$B>0$ to be smaller than the presently measured speed of light $c_0$.

The laws of physics will be interpreted differently by observers, depending upon whether
they perform experiments in the VSL frame or the VSGW frame.

The Friedmann equation in the VSL frame is given by
\begin{equation}
\label{Friedmann}
H^2+\frac{kc_0^2}{R^2} = \frac{8\pi G}{3I^{1/2}}\rho
+\frac{1}{3}c_0^2\Lambda+\frac{1}{6}\rho_\phi,
\end{equation}
where
\begin{equation}
\rho_\phi=\frac{1}{2}\dot\phi^2+c_0^2V(\phi).
\end{equation}
The remaining equation is
\begin{equation}
\label{eq:ddot eqn}
\frac{{\ddot R}}{R}=-\frac{4\pi
G}{3I^{1/2}}\biggl(\rho+3I\frac{p}{c_0^2}\biggr)
+\frac{1}{3}c_0^2\Lambda-\frac{1}{12}(\rho_\phi +3p_\phi),
\end{equation}
where
\begin{equation}
p_\phi=\frac{1}{2}\dot\phi^2-c_0^2V(\phi).
\end{equation}

The scalar field equation (\ref{eq:scalar FEQ}) in the VSL frame is
\begin{equation}
\label{eq:phi dot}
\frac{1}{c_0^2}\biggl(1-\frac{16\pi
GB}{c_0^2I^{3/2}}\rho\biggr)\ddot{\phi}+\frac{3}{c_0^2}H\dot{\phi}
\biggl(1+\frac{16\pi GB}{c_0^4I^{1/2}}p\biggr)+V^\prime(\phi)=0.
\end{equation}

The Friedmann equation in the VSGW frame is given by
\begin{equation}
\label{Kfriedmann}
H^2+\frac{c_0^2kK}{R^2}=\frac{8\pi G}{3}K^{3/2}\rho
+\frac{1}{3}c_0^2\Lambda K+\frac{1}{6}{\tilde\rho}_\phi.
\end{equation}
where we have defined
\begin{equation}
{\tilde\rho}_\phi=\frac{1}{2}\dot\phi^2+c_0^2KV(\phi),\quad {\tilde p}_\phi
=\frac{1}{2}\dot\phi^2-c_0^2KV(\phi).
\end{equation}
The remaining equation is
\begin{equation}
\label{Kacceleration}
\frac{\ddot R}{R}=-\frac{4\pi G}{3}\biggl(K^{3/2}\rho+\frac{3}{c_0^2}K^{1/2}p\biggr)
+\frac{1}{3}c_0^2\Lambda K-\frac{1}{12}({\tilde\rho}_\phi+3{\tilde p}_\phi)
+\frac{1}{2}\frac{\dot K}{K}H.
\end{equation}

The scalar wave equation in this frame is given by
\begin{equation}
\label{GWscalar}
\frac{1}{c_0^2}\biggl(1-\frac{16\pi GB}{c_0^2}K^{3/2}\rho\biggr)\ddot\phi
+\frac{3K}{c_0^2}H\dot\phi\biggl(1+\frac{16\pi GB}{c_0^4}K^{1/2}p\biggr)
+K^2V^\prime(\phi)=0.
\end{equation}

\section{Calculation of the Red Shift}

By adopting the VSL frame and setting $d{\hat s}^2=0$ in
(\ref{vslmetric}), we obtain for fixed $\theta$ and $\phi$:
\begin{equation}
\label{cintegral}
\int_{t_1}^{t_0}\frac{dt c(t)}{R(t)}=\int_0^{r_1}\frac{dr}{\sqrt{1-kr^2}}=f(r_1)
={\rm const.},
\end{equation}
where $t_1$ denotes the time a light wave leaves a supernova or galaxy and $t_0$
denotes the time when it reaches us on earth. If the next wave crest leaves $r_1$ at
time $t_1+\delta t_1$, it will arrive on earth at time $t_0+\delta t_0$ giving
\begin{equation}
\label{int2}
\int^{t_0+\delta t_0}_{t_1+\delta t_1}\frac{dtc(t)}{R(t)}=f(r_1).
\end{equation}
Subtracting (\ref{cintegral}) from (\ref{int2}) and considering that $R$
and $c$ vary little during the period of a light signal, we obtain
\begin{equation}
\frac{c(t_0)\delta t_0}{R(t_0)}=\frac{c(t_1)\delta t_1}{R(t_1)}.
\end{equation}
The frequency at emission $\nu_1$ is then related to the observed frequency $\nu_0$
by
\begin{equation}
\frac{\nu_0}{\nu_1}=\frac{\delta t_1}
{\delta t_0}=\frac{c(t_0)}{c(t_1)}\frac{R(t_1)}{R(t_0)}.
\end{equation}

The red shift
\begin{equation}
z=\frac{\lambda_0-\lambda_1}{\lambda_1}
\end{equation}
is then given by
\begin{equation}
\label{redshift}
z=\frac{c(t_1)}{c(t_0)}\frac{R(t_0)}{R(t_1)}-1,
\end{equation}
where we have
\begin{equation}
\lambda(t)\nu(t)=c_0,
\end{equation}
so that $\lambda_0/\lambda_1=\nu_1/\nu_0$. This follows from the relation
\begin{equation}
{\hat k}^\mu{\hat k}_\mu={\hat g}^{\mu\alpha}k_\alpha{\hat
g}_{\mu\beta}k^\beta=k_\beta k^\beta=0,
\end{equation}
where we have used (\ref{kronecker}) and $k^\mu=(\nu/c_0,1/{\vec\lambda})$ and
$\lambda=\vert{\vec\lambda}\vert$.

We see that if $c(t_1) > c(t_0)$, then the observed red shift will appear to be
larger due to the speed of light in the past being bigger than the presently observed
speed $c(t_0)=c_0$ in an expanding universe with $R(t_0) > R(t_1)$.

\section{Apparent Luminosity and Luminosity Distance}

The apparent luminosity $l$ is the power per unit mirror area in our VSL
frame~\cite{Weinberg}:
\begin{equation}
l\equiv
\frac{P}{A}=L\biggl(\frac{c^2(t_0)}{c^2(t_1)}\biggr)\biggl(\frac{R^2(t_1)}{R^2(t_0)}\biggr)
\biggl(\frac{1}{4\pi R^2(t_0)r_1^2}\biggr),
\end{equation} where $L$ is the absolute
luminosity, $A$ is the proper area of the mirror and $P$ is the total power emitted
at the source.
The luminosity distance of a light source in Euclidean space is
\begin{equation}
d_L=\biggl(\frac{L}{4\pi l}\biggr)^{1/2},
\end{equation}
which leads to the expression
\begin{equation}
\label{lumdistance}
d_L=\biggl(\frac{c(t_1)}{c(t_0)}\biggr)\biggl(\frac{R^2(t_0)r_1}{R(t_1)}\biggr).
\end{equation}

Let us expand $R(t)$ as
\begin{equation}
\label{Rexpansion}
R(t)=R(t_0)[1+H_0(t-t_0)-\frac{1}{2}q_0H_0^2(t-t_0)^2+...],
\end{equation}
where
\begin{equation}
H_0=\frac{{\dot R}(t_0)}{R(t_0)},\quad q_0=-\frac{{\ddot R}(t_0)R(t_0)}{{\dot
R}^2(t_0)}.
\end{equation}
We also expand $c(t)$ in the power series
\begin{equation}
\label{cexpansion}
c(t)=c(t_0)[1+D_0(t-t_0)+\frac{1}{2}Q_0D_0^2(t-t_0)^2+...],
\end{equation}
where
\begin{equation}
D_0=-\frac{{\dot c}(t_0)}{c(t_0)},
\quad Q_0=-\frac{{\ddot c}(t_0){c(t_0)}}{{\dot c}^2(t_0)}.
\end{equation}
The negative signs of the coefficients $D_0$ and $Q_0$ are chosen to
yield a decreasing speed of light as the universe expands. From
(\ref{redshift}), we get using (\ref{Rexpansion}) and (\ref{cexpansion})
the red shift
\begin{equation}
z=(H_0-D_0)(t_0-t_1)+[(1+\frac{1}{2}q_0)H_0^2+\frac{1}{2}Q_0D_0^2-D_0H_0](t_0-t_1)^2+...
\end{equation}
Inverting this equation we get
\begin{equation}
t_0-t_1=\biggl(\frac{1}{H_0-D_0}\biggr)z-\biggl(\frac{1}{H_0-D_0}\biggr)^3
[(1+\frac{1}{2}q_0)H_0^2+\frac{1}{2}Q_0D_0^2-D_0H_0]z^2+....
\end{equation}

From (\ref{cintegral}), we obtain
\begin{equation}
\biggl(\frac{c(t_0)}{R(t_0)}\biggr)[t_0-t_1+\frac{1}{2}(H_0-D_0)(t_0-t_1)^2+..]
=r_1+O(r_1^3).
\end{equation}
This gives the result for $r_1$:
\begin{equation}
r_1=\biggl(\frac{c(t_0)}{R(t_0)}\biggr)
\biggl\{\biggl(\frac{1}{H_0-D_0}\biggr)z
-\biggl(\frac{1}{H_0-D_0}\biggr)^3[(1+\frac{1}{2}q_0)H_0^2+\frac{1}{2}Q_0D_0^2-D_0H_0]z^2
$$ $$
+\frac{1}{2}\biggl(\frac{1}{H_0-D_0}\biggr)z^2+...\biggr\}.
\end{equation}
From (\ref{redshift}) and (\ref{lumdistance}), we find
\begin{equation}
\label{luminositydist}
d_L=c(t_0)r_1(1+z)
$$ $$
=\biggl(\frac{c(t_0)}{H_0-D_0}\biggr)\biggl\{z-\biggl(\frac{1}{H_0-D_0}\biggr)^2
[(1+\frac{1}{2}q_0)H_0^2+\frac{1}{2}Q_0D_0^2-D_0H_0
$$ $$
-\frac{3}{2}(H_0-D_0)^2]z^2+...\biggr\}.
\end{equation}
In the limit that
$D_0=Q_0=0$, we obtain the correct GR limit
\begin{equation}
d_L=\biggl(\frac{c(t_0)}{H_0}\biggr)[z+\frac{1}{2}(1-q_0)z^2+...].
\end{equation}

We now obtain the formula for the apparent luminosity
\begin{equation}
l\equiv\frac{L}{4\pi d_L^2}
=\biggl(\frac{L}{4\pi
z^2c^2(t_0)}\biggr)(H_0-D_0)^2\biggl\{1+\biggl(\frac{1}{H_0-D_0}\biggr)^2
[(1+\frac{1}{2}q_0)H_0^2
$$ $$
+\frac{1}{2}Q_0D_0^2-D_0H_0-\frac{3}{2}(H_0-D_0)^2]z^2+...\biggr\}.
\end{equation}

\section{Dimming of Supernovae}

Let us write the Friedmann equation (\ref{Friedmann}) as
\begin{equation}
\label{Omegasum}
\Omega_{0m}+\Omega_{0k}+\Omega_{0\Lambda}+\Omega_{0\phi}=1,
\end{equation}
where
\begin{equation}
\Omega_{0m}=\frac{8\pi G\rho_{0m}}{3H^2_0I_0^{1/2}},\quad
\Omega_{0k}=-\frac{c^2_0 k}{R_0^2H_0^2},\quad
\Omega_{0\Lambda}=\frac{c_0^2\Lambda}{3H_0^2},
$$ $$
\Omega_{0\phi}=\frac{\rho_{0\phi}}{6H_0^2},
\end{equation}
where $\Omega_0, \rho_0, \phi_0$ and $H_0$ denote
the present values of the corresponding quantities.

Choosing the values
\begin{equation}
\Omega_{0m}=0.28,\quad \Omega_{0k}=0,\quad \Omega_{0\Lambda}=0,\quad
\Omega_{0\phi}=0.72,
\end{equation}
we obtain a fit to the supernovae data
with $q_0=\frac{1}{2}$ and $D_0$ and $Q_0$
chosen for a given red shift $z$ to give a 10-15\% increase in the luminosity distance
(\ref{luminositydist}), corresponding to a 20-30\% decrease in the apparent luminosity
observed in the supernovae data~\cite{Perlmutter}. For a negligible value
of the pressure $p_m$, we get from (\ref{eq:ddot eqn}) with $\Lambda=0$:
\begin{equation}
\frac{{\ddot R}}{R}=-\biggl[\biggl(\frac{4\pi
G}{3I_0^{1/2}}\biggr)\rho_{0m}+\frac{1}{12}(\rho_{0\phi}+3p_{0\phi})\biggr].
\end{equation}
We see that
the scalar field $\phi$ does not produce an acceleration of the universe,
when $\rho_{0\phi}>0$ and $p_{0\phi}>0$. The increased speed of light at the
observed SNe Ia red shifts $z$ has dimmed the supernovae in accordance with
the SNe Ia observations. We do not attempt here to explain why $\Lambda=0$.

The Friedmann equation (\ref{Friedmann}) can be rewritten in terms of an effective
gravitational constant $G_{\rm eff}=G/\sqrt{I(t)}$, corresponding to a time-varying
gravitational constant.

We see that the time variation of $G$ depends on
the time derivative of the scalar field $\phi$. The cosmic time evolution
of $G$ should be of the order of the expansion rate of the universe, i.e.
${\dot G}/G \sim H_0$, where $H_0=100h\, {\rm
km}\,s^{-1}\,{\rm Mpc}^{-1}=h\times 10^{-10}\,{\rm yr}^{-1}$~\cite{Will}.
Current observations of the expansion of the universe yield $h\sim 0.7$.
For $z\sim 1$ an estimate gives $\Delta t\sim 4.9\times 10^9\, {\rm
yr}$, so that we obtain $\Delta G/G\sim 0.3$. Thus, in the red shift
range $z\sim
0.1-3$, a 10-15\% change in $G$ can be accomodated and be consistent with a
10-15\% change in $c(t)
=c_0(1+\frac{B}{c_0^2}{\dot\phi}^2)^{1/2}$. However, as we approach the
time of big-bang nucleosynthesis (BBN) at a red shift $z\sim
10^9-10^{10}$, we have $\Delta G/G\sim 10^{-9}-10^{-10}$ and the variation
of $G$ with time becomes much more restrictive. This requires that for
red shifts larger than $z\sim 3$, the scalar field $\phi$ should tend to
a constant and $I(t)\rightarrow 1$ in order not to violate the good
agreement of BBN calculations. A more detailed numerical analysis of
solutions of the BGT for $z\sim 10^9$ is needed to determine the behavior
of $I(t)$. Such an analysis will be considered in a future publication.

If we now view the expansion of the universe from within the gravitational metric or
VSGW frame, then the speed of light is constant but gravitational waves will move
with a speed different from $c_0$. From (\ref{Kfriedmann}), we obtain the
contributions to (\ref{Omegasum}):
\begin{equation}
\Omega_{0m}=\frac{8\pi GK_0^{3/2}\rho_{0m}}{3H_0^2},\quad
\Omega_{0k}=-\frac{c_0^2kK_0}{R_0^2H_0^2},\quad
\Omega_{0\Lambda}=\frac{c_0^2K_0\Lambda}{3H_0^2},
$$ $$
\Omega_{0\phi}=\frac{{\tilde\rho}_{0\phi}}{6H_0^2}.
\end{equation}
We obtain from (\ref{Kfriedmann}) the deceleration parameter
\begin{equation}
q=\frac{dH^{-1}}{dt}-1=-\frac{{\dot K}}{2HK}+\frac{1}{2}(1+\Omega_k)
+\frac{{\tilde p}_\phi}{4H^2}+\frac{4\pi
G}{c_0^2H^2}K^{1/2}p_m-\frac{c_0^2K\Lambda}{2H^2}.
\end{equation}
By using the scalar wave equation (\ref{GWscalar}), we find
that~\cite{ClaytonMoffat2,ClaytonMoffat4}:
\begin{equation}
\frac{{\dot K}}{2HK}=\frac{3(1-K)(1+16\pi GBK^{1/2}p_m)+H^{-1}B\dot\phi
KV^\prime(\phi)}
{1-16\pi GBK^{3/2}\rho_m/c_0^2}.
\end{equation}
The first term in the numerator and the denominator are positive-definite whenever
the metrics have the correct signature, and the second term in the numerator is
positive provided that $H > 0$ and $\dot\phi KV^\prime(\phi) > 0$.

If we choose $p_m\sim 0$ and $k=\Lambda=0$, we get a fit to the current data
in the VSGW frame with
\begin{equation}
\Omega_{0m}=0.28,\quad \Omega_{0\phi}=0.72,\quad q_0=-0.63,
\end{equation}
where
\begin{equation}
\label{qacceleration}
q_0=-\frac{{\dot K_0}}{2H_0K_0}+\frac{1}{2}+\frac{{\tilde p}_{0\phi}}{4H_0^2}.
\end{equation}
We can achieve an acceleration of the universe, if the first term on the right-hand
side of (\ref{qacceleration}) dominates, which means that the cosmic acceleration
with ${\ddot R}>0$ is caused by the dynamics of the scalar field $\phi$ with
$\Lambda=0$ in the VSGW frame. Provided that $q$ turns small and positive
at an earlier time in the universe's expansion, then galaxy formation can
be achieved.

Thus, in the VSL frame the universe appears to be decelerating and the
supernovae are dimmed because of the increase of the speed of light with increasing
red shift $z$ at least up to $z\sim 3-4$, while in the VSGW frame the universe
appears to be accelerating and the supernovae appear to be farther away.

We have demonstrated elsewhere, that the early universe standard horizon and
flatness problems can be resolved either in the VSL frame with a varying speed of
light, or within the VSGW frame with constant speed of light, but with a varying
speed of gravitational
waves~\cite{Moffat,Moffat2,ClaytonMoffat,ClaytonMoffat2,ClaytonMoffat3,ClaytonMoffat4}.

\section{Fine-Structure Constant and Energy Conservation}

Let us consider the behavior of the fine-structure constant $\alpha=e^2/\hbar
c$ and the energy of a system of particles in our BGT.
We shall adopt the action for a charged particle moving in an electromagnetic field
\begin{equation}
\label{emaction}
S_{EM}=-\int d{\hat s}\biggl[mc_0\biggl({\hat u}^\mu
{\hat u}^\nu{\hat g}_{\mu\nu}\biggr)^{1/2}
+\frac{e_0}{c_0}{\hat u}^\mu{\hat A}_\mu\biggr],
\end{equation}
where
\begin{equation}
F_{\mu\nu}=\partial_\mu A_\nu-\partial_\nu A_\mu,
\end{equation}
is the electromagnetic field. Moreover, ${\hat F}^{\mu\nu}={\hat g}^{\mu\alpha}{\hat
g}^{\nu\beta}F_{\alpha\beta}$ is the electromagnetic field observed in the matter
VSL frame, ${\hat u}^\mu={\hat g}^{\mu\alpha}u_\alpha$, ${\hat A}_\mu={\hat
g}_{\mu\nu}A^\nu$, $m$ is the particle mass and $e_0$ is the constant electron
charge.  We have in our isotropic and homogeneous cosmology
$A^\mu=(e_0/r,0)$.

From (\ref{kronecker}) and (\ref{emaction}), we obtain for the
Coulomb energy
\begin{equation}
V_{\rm Coul}
=-\frac{e^2_0}{c_0r}\frac{dx_0}{ds},
\end{equation}
where $dx_0/ds=\gamma c_0$ with $\gamma=1/(1-u^2/c_0^2)$. We obtain in the
nonrelativistic limit
\begin{equation}
V_{\rm Coul}=-\frac{e^2_0}{r}.
\end{equation}
In particular, the fine-structure constant
\begin{equation}
\alpha\equiv \frac{e^2}{\hbar c}=\frac{e_0^2}{\hbar c_0}
\end{equation}
is constant in our bimetric gravity theory. This will be true whether or not we
observe $\alpha$ in the VSL frame or in the VSGW frame.

This does not conform with the observations of absorption line spectra from
quasars~\cite{Webb}, which yield
\begin{equation}
\label{alpha}
\frac{\Delta\alpha}{\alpha}=-0.72\pm 0.18\times 10^{-5}.
\end{equation}
However, these data still require to be confirmed by an independent measurement
procedure before we can rule out the constancy of $\alpha$ predicted by the BGT.
If we wish to interpret the dimming of the supernovae by a $\sim 10\%$ increase
in the speed of light in the VSL frame, then we cannot simultaneously demand a
similar decrease in the fine-structure constant $\alpha$, because this would strongly
disagree with the Webb et al. result (\ref{alpha}) and lead to serious
violations of the weak equivalence principle experimental tests. A recent analysis of
the effects of a time varying $\alpha$ in the CMB background, gives results which
are consistent with no variation in $\alpha$ from the epoch of recombination to the
present day, and restricts any such variation to be less than about 4\%~\cite{Martins}.
Forthcoming MAP and Planck experiments will be able to measure variations in $\alpha$
to better than a percent accuracy.

Finally, let us consider the conservation of energy in BGT. We have from
(\ref{kronecker}):
\begin{equation}
{\hat p}^\mu{\hat p}_\mu={\hat g}^{\mu\alpha}p_\alpha{\hat g}_{\mu\beta}p^\beta
=p^\mu p_\mu=(mc_0)^2.
\end{equation}
From the identification $p^\mu=(E/c_0,{\vec p})$, we obtain the standard special
relativity result
\begin{equation}
E=c_0[({\vec p})^2+(mc_0)^2]^{1/2}.
\end{equation}

\section{Cosmological Horizons}

According to the models of
quintessence~\cite{Peebles}, the dark energy of the universe is dominated by the
potential $V(\phi)$ of a scalar field $\phi$, which rolls down to its minimum at
$V=0$. We recall that for an equation of state $p=w\rho$ in GR, a
cosmological constant corresponds to $w=-1$, radiation domination to
$w=\frac{1}{3}$ and matter domination to $w=0$. On the other hand,
quintessence gives an equation of state with
\begin{equation}
\label{wrange} -1 < w < -\frac{1}{3},
\end{equation} while
the observational evidence for a cosmological constant is given by the
bound:
\begin{equation}
-1 < w_{\rm observed} \leq -\frac{2}{3}.
\end{equation}

In the VSL frame, the proper horizon distance is given by
\begin{equation}
\delta_H(t)=R(t)F,
\end{equation}
where
\begin{equation}
\label{Fintegral}
F=\int^\infty_{t_0}\frac{dt^\prime c(t^\prime)}{R(t^\prime)}.
\end{equation}
Whenever $F$ diverges there exist no future event horizons
in the spacetime geometry. On the other hand, when $F$ converges the
spacetime geometry exhibits a future horizon, and events whose coordinates
at time ${\bar t}$ are located beyond $\delta_H$ can never communicate with
the observer at $r=0$.

The variation of the expansion scale factor at large $R(t)$, when the
curvature becomes negligible, approaches
\begin{equation}
R(t)\sim t^{2/3(1+w)}.
\end{equation}
We now have
\begin{equation}
\label{Fint}
F=c_0\int^\infty_{t_0}dt^\prime t^{\prime[2(n-1)/3(1+w)]},
\end{equation}
where we have assumed that a solution of the BGT field equations leads to the
behavior
\begin{equation}
c(t)\sim c_0R^n,
\end{equation}
where $n$ is some positive number. In the VSL frame with the metric ${\hat
g}_{\mu\nu}$, we can choose $w>1/3$, $\Lambda=0$ and a value of $n$ such that the
integral $F$ diverges and there is no future cosmological horizon. On the other
hand, in the VSGW frame with the metric $g_{\mu\nu}$, the universe is accelerating
with ${\ddot R}>0$, the velocity of light is constant, $c(t)=c_0$, and we expect
that the integral $F$ will converge and allow a future cosmological horizon.

In an eternally accelerating universe, we are confronted with the difficulty
of defining a consistent S-matrix description of quantum field theory and
string/M-theories. We have shown that if we restrict ourselves to the VSL
frame with a varying speed of light and a decelerating universe, then we can
avoid asymptotically all future horizons associated with quintessence
models and an accelerating universe. On the other hand, if we perform our
experiments and theoretical calculations in the VSGW frame, then we can
expect to encounter future cosmological horizons. Thus, it is advantageous
for us to restrict ourselves to the VSL frame, since we can construct a
viable S-matrix for quantum field theories and string/M-theories with a
satisfactory asymptotic null infinity.

\section{Conclusions}

The problem of constructing a self-consistent gravity theory which is
diffeomorphism invariant and allows a varying speed of light and a varying speed
of gravitational waves can be solved within the BGT formalism. A proper conservation
law for the matter tensor can be constructed that leads to particles moving on
geodesics in the comoving ${\hat g}_{\mu\nu}$ frame. In this frame, the speed of
light is constant and observers can communicate with one another by light signals.
An observer in the spacetime determined by the $g_{\mu\nu}$ metric, sees the speed
of gravitational waves varying with time and the universe accelerate and a future
cosmological horizon is expected to exist.

On the other hand, when the metric $g_{\mu\nu}$ is chosen
to be comoving, so that the speed of gravitational waves is constant, and observers
will be able (in the future) to communicate with one another by means of
gravitational wave signals, then an observer in the ${\hat g}_{\mu\nu}$ frame will
detect a varying speed of light, a dimming of supernovae because of the
increase of $c(t)$ in the past universe, and a decelerating universe
without a cosmological horizon for an increasing speed of light in the future
universe.

The observer in the VSL frame will be in an advantageous position when performing
physical experiments, because the non-existence of a future cosmological horizon will
permit the construction of a meaningful S-matrix for the scattering of particles.

An important feature of the BGT is that the dimensionless fine-structure constant
$\alpha$ is constant in spacetime, so problems associated with possible violations of
the weak equivalence principle tests and massive fine-tuning of vacuum
quantum field theory can be avoided~\cite{Banks}. However, future
experiments will hopefully decide whether the putative observations of a
variation of $\alpha$ by analyses of quasar absorption line spectra are
valid~\cite{Webb}.

There are many interesting consequences of BGT that can be explored e.g. the behavior
of collapsing stars and the nature of dispersion relations at high energy. It is
also of interest to investigate the possible role of BGT in quantum
gravity. These topics will be the subject of future invetigations.

\vskip 0.2 true in {\bf Acknowledgments}
\vskip 0.2 true in I thank Michael Clayton for helpful and
stimulating discussions. This work was supported by the Natural Sciences and
Engineering Research Council of Canada.  \vskip 0.5 true in

\end{document}